\documentclass[pre,twocolumn,showpacs,superscriptaddress]{revtex4}
\usepackage{graphicx,amsmath,bm}

\newcommand{\pdiff}[3][]{\frac{\partial^{#1} #2}{\partial #3^{#1}}}     
\renewcommand*\vec[1]{\bm #1}                                           
\begin{document}

%......Title and other stuff 
\title{Polarons in semiconductor quantum-dots and their role in the
       quantum kinetics of carrier relaxation}

\author{J. Seebeck} 
\author{T.R. Nielsen} 
\affiliation{Institute for Theoretical Physics,  
             University of Bremen,
             28334 Bremen, Germany}

\author{P. Gartner}
\affiliation{Institute for Theoretical Physics,  
             University of Bremen,
             28334 Bremen, Germany}
\affiliation{National Institute for Materials Physics, POB MG-7, 
             Bucharest-Magurele, Romania}

\author{F. Jahnke}
%\email[Corresponding author: ]{www.itp.uni-bremen.de/~jahnke}
\affiliation{Institute for Theoretical Physics,  
             University of Bremen,
             28334 Bremen, Germany}

\date{\today}


\begin{abstract}
  While time-dependent perturbation theory shows inefficient
  carrier-phonon scattering in semiconductor quantum dots, we
  demonstrate that a quantum kinetic description of carrier-phonon
  interaction predicts fast carrier capture and relaxation. The
  considered processes do not fulfill energy conservation in terms of
  free-carrier energies because polar coupling of localized quantum-dot
  states strongly modifies this picture.
\end{abstract}
%......end of Title and other stuff 

\maketitle

\section{INTRODUCTION}

Applications of semiconductor quantum dots (QDs) in optoelectronic
devices rely on fast carrier scattering processes towards and between
the discrete confined levels. These carrier transitions determine the
dynamics of QD luminescence \cite{Morris:99} or the operation of QD
lasers \cite{Bhattacharya:99,Deppe:00}.  For low carrier densities,
where Coulomb scattering can be neglected, carrier-phonon interaction
provides the dominant scattering channel. In QDs only phonons with
small momenta can efficiently couple to the confined carriers
\cite{Inoshita:97}.  Then interaction with LA phonons does not
contribute for large transition energies and only quasi-monochromatic
LO phonons need to be considered.

The simplest theoretical approach to electronic scattering processes
is based on time-dependent perturbation theory. Fermi's golden rule
for carrier transitions due to phonon emission or absorption contains a
delta-function for strict energy conservation in terms of
free-carrier energies of initial and final states and the phonon
energy.  When transition energies of localized QD states do not
match the LO-phonon energy, efficient scattering is inhibited (leading
to the prediction of a phonon bottleneck) and only higher-order
processes, like a combination of LO and LA phonons
\cite{Inoshita:92,Singh:98}, weakly contribute.  Attempts to broaden
the delta-function ``by hand'' immediately change the results
\cite{Singh:98} which underlines that this point should be addressed
microscopically.
The phonon bottleneck effect is still a debated topic,
with experimental evidence both for \cite{Urayama:01,Minnaert:01,Xu:02} 
and against it \cite{Tsitsishvili:02,Peronne:03,Quochi:03}.

As in any coupled system, carrier-phonon interaction renormalizes both
electronic and vibrational states. However, in bulk semiconductors or
quantum wells with weak polar coupling, the net effect can be
described by renormalized effective carrier masses, a small polaron
shift of the band-edge, and lattice distortions only modify the
background dielectric constant for the Coulomb interaction of
carriers. The broadening of the transition energies due to
carrier-phonon interaction remains weak.

For carriers in QDs, the discrete nature of localized electronic
states changes the role of polaronic effects
\cite{Inoshita:97,Kral:98,Verzelen:02}.  Restricting the analysis to a
single QD state coupled to phonons, polaron effects can be obtained
from an exact diagonalization of the Hamiltonian \cite{Mahan:90}.
While an extension to several discrete levels has been presented
\cite{Stauber:00}, the influence of the energetically nearby continuum
of wetting-layer (WL) states, typical for self-assembled QDs, has not
been included.  Furthermore, only quasiparticle properties have been
discussed which provide no direct information about the scattering
efficiency for various processes.  Calculations of carrier transition
rates based on the polaron picture are missing.

We use a quantum kinetic treatment for carrier-phonon interaction in
the polaron picture. As a first step, quasi-particle renormalizations
due to the polar interaction for both QD and WL carriers are
determined.  For the QD states, the hybridization of one state with
strong satellites of another state leads to a rich multi-peak
structure.  The WL states exhibit weak LO-phonon satellites. Coupling
to the WL states provides a broadening mechanism for the QD states.

Based on the spectral properties of QD and WL polarons, quantum
kinetic equations for the capture process (carrier transitions from
the WL into the QD) and relaxation processes (transitions between QD
states) are solved.  For situations where, in terms of free-carrier
energies, energy conserving scattering processes are not possible, the
quantum-kinetic treatment provides efficient scattering rates.  Even for
the InGaAs material system with weak polar coupling, sub-picosecond
scattering times are obtained.

%%%%%%%%%%%%%%%%%%%%%%%%%%%%%%%%%%%%%%%%%%%%%%%%%%%%%%%%%%%%%%%%%%%%%%%%%%
\section{Quantum dot polarons}

The single-particle properties of carriers under the influence of
lattice distortions are determined by the retarded Green's function
(GF), $G^r_{\alpha}$, which obeys the Dyson equation
%%%%%%%%%%%%%%%%
\begin{multline}
   \Big[ i\hbar\pdiff{}{t_1} - \varepsilon_{\alpha} \Big]
     ~G^r_{\alpha}(t_1,t_2)
   = \delta(t_1-t_2) \\
   + \int\!dt_3 ~~ \Sigma^r_{\alpha}(t_1,t_3) ~ G^r_{\alpha}(t_3,t_2).
\label{eq:Dyson_G_ret}
\end{multline}
%%%%%%%%%%%%%%%
Here $\alpha$ is an arbitrary (QD or WL) electronic
state with free-carrier energy $\varepsilon_{\alpha}$.  In the polaron
theory one usually considers all possible virtual transitions from
this state due to emission or absorption of phonons.  This corresponds
to a self-energy $\Sigma^r_{\alpha}$ for the carrier-phonon
interaction where the population of the involved carrier states is
neglected (electron vacuum). The corresponding retarded self-energy 
in random-phase approximation (RPA) is given by \cite{note:diag_GF}
%%%%%%%%%%%%%%%%
\begin{equation}
      \Sigma^r_{\alpha}(t_1,t_2) = i \hbar \sum_{\beta} ~
      G^r_{\beta}(t_1,t_2) ~ D^<_{\beta \alpha}(t_2-t_1).
\label{eq:Sigma_ret}
\end{equation}
%%%%%%%%%%%%%%%
Assuming that the phonon system is in thermal equilibrium,
the phonon propagator (combined with the interaction matrix elements) 
is given by
%%%%%%%%%%%%%%%%
\begin{multline}
   i\hbar ~ D^<_{\beta \alpha}(\tau)=\sum_{\vec{q}} ~
   |M_{\beta \alpha}(\vec{q})|^2 \\ \times
   \Big[    n_{LO}  ~ e^{-i\omega_{LO}\tau}
      ~+~(1+n_{LO}) ~ e^{ i\omega_{LO}\tau} \Big]
\label{eq:pn_propagator}
\end{multline}
%%%%%%%%%%%%%%%
where monochromatic LO-phonons with the frequency $\omega_{LO}$ are
considered. The corresponding phonon population is given by
$n_{LO}=1/(e^{\hbar\omega_{LO}/kT}-1)$ and the Fr\"ohlich interaction
matrix element
%%%%%%%%%%%%%%%%
\begin{equation}
  M_{\beta \alpha}(\vec{q})=\frac{ M_{LO} }{ q }~
  \langle \beta | e^{i\vec{q}\vec{r}} | \alpha\rangle 
\label{eq:coupling}
\end{equation}
%%%%%%%%%%%%%%%
contains the overlap between the electronic states and the phonon
mode. For localized electronic states this acts as a form factor.  The
prefactor $M^2_{LO}=4\pi\alpha \frac{\hbar}{\sqrt{2m}}
(\hbar\omega_{LO})^{3/2}$ includes the polar coupling strength
$\alpha$ and the reduced mass $m$.  As a result of the above
assumptions, the retarded GF itself depends only on the difference of
time arguments and its Fourier transform can be directly related to
the quasi-particle properties.

Due to energy separation between the discrete QD states and the WL
continuum, polaronic effects in QDs are often computed by neglecting
the presence of the WL
\cite{Inoshita:97,Kral:98,Verzelen:02,Stauber:00}.  For a single
discrete level this amounts to the exactly solvable independent Boson
model \cite{Mahan:90} and for several discrete levels it was shown to
be nearly exactly solvable \cite{Stauber:00}. In both cases, even for
non-zero temperatures, the spectral function contains a series of
sharp delta-like peaks. In real QDs, however, the interaction with the
WL continuum (which might require multi-phonon processes) leads to a
broadening of these peaks. The RPA accounts for this broadening effect
while it retains a hybridization effect (see below) characteristic for
the full solution. Therefore the RPA is expected to provide an
adequate description {\it in the presence of the continuum}
\cite{note:RPA}.  An additional source of broadening is the finite
LO-phonon lifetime due to anharmonic interaction between phonons.

For the numerical results presented in this paper we consider an
InGaAs QD-WL system with weak polar coupling $\alpha=0.06$.  The
effective-mass approximation is assumed to be valid with $m_e=0.067
m_0$ for the conduction band.  For flat lens-shaped QDs the in-plane
wave-functions of an isotropic two-dimensional harmonic potential are
used while for the (strong) confinement in the direction perpendicular
to the WL a finite-height potential barrier is considered (see
\cite{Nielsen:04} for parameters and further details).  To account for
a finite height of the QD confinement potential, the calculations only
include the (double degenerate) ground state and the (four-fold
degenerate) first excited state, in the following called s and
p-shell, respectively with s-p spacing and p-WL separation of 40 meV.
For the description of the WL states we use the following steps: (i)
the WL states in the absence of QDs are described by plane waves for
the in-plane part, multiplied by the state corresponding to the
finite-height barrier confinement for the perpendicular direction.
(ii) to describe the WL states in the presence of the QDs, the
orthogonalized plane wave scheme, described in Appendix A of
Ref.~\cite{Nielsen:04}, is used to construct WL states orthogonal to
the QD states. Calculations are done for a density of QDs on the WL
$n_{dot}= 10^{10}$ cm$^{-2}$. Details on the calculation of the
interaction matrix elements in Eq.~(\ref{eq:coupling}) with these wave
functions for various combinations of QD and WL states can be found in
Appendix B of Ref.~\cite{Nielsen:04}. Convergent results are obtained
with 128 points for the in-plane momentum radial integrals and 50 points for
the remaining angular integrations entering the interaction matrix
elements.

%%%%%%%%%%%%%%%%%%%%%%%%%%%%%%%%%%%%%%%%%%%%%%%%%%%%%%%%%%%%%%%%%%%%%%%%
\begin{figure}[htb]
\includegraphics*[width=.45\textwidth]{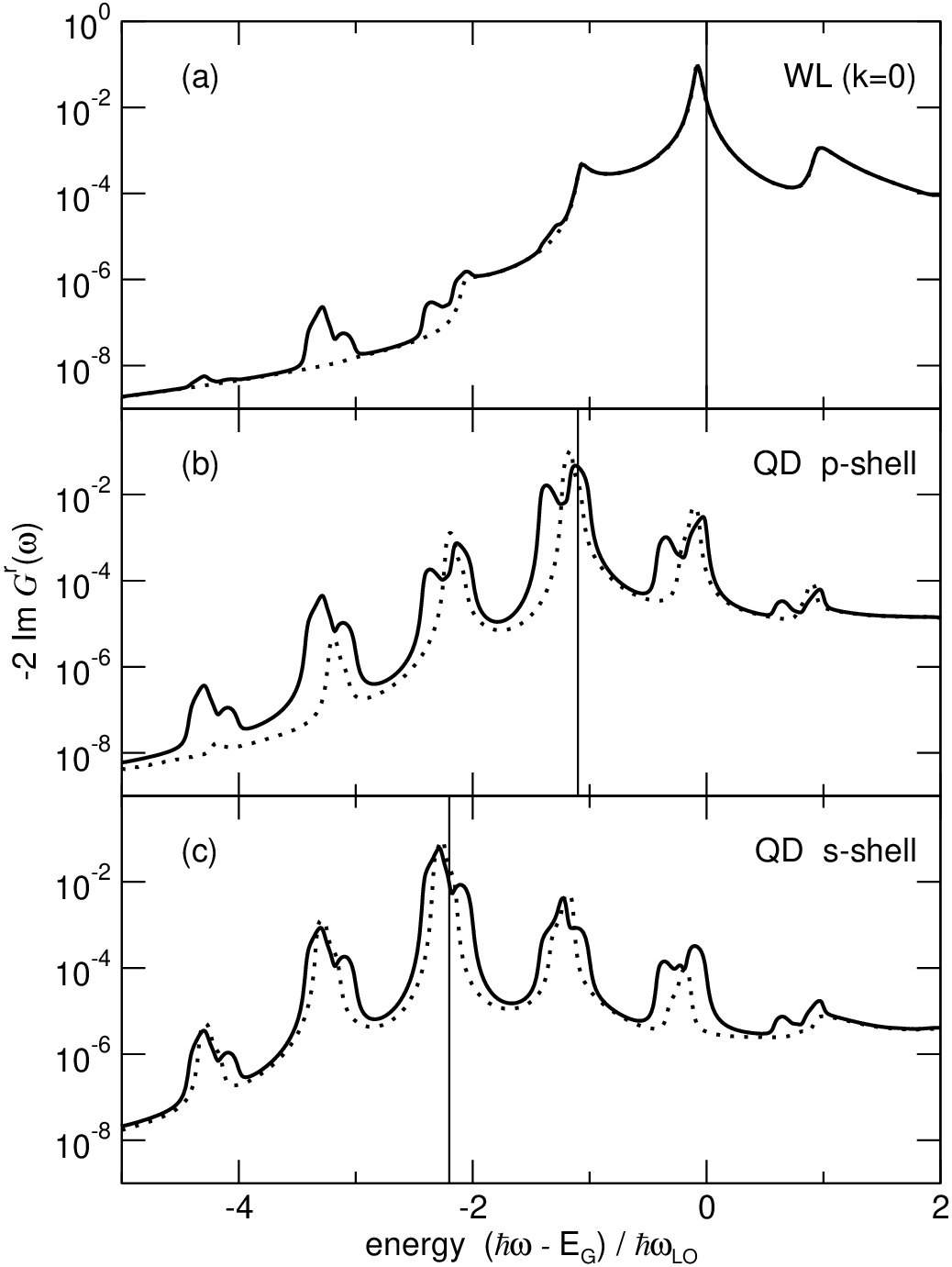}
\caption{(a) Spectral function of electrons for the lowest WL state
  at $k=0$ under the influence of polar coupling in the combined QD-WL
  system (solid line) and without coupling to the QD states (dotted
  line).  For electrons in the QD p-shell (b) and s-shell (c)
  full coupling between all states (solid line) is compared to the case
  without coupling to other QD states (dotted line). Vertical lines
  show the corresponding free-carrier energies.  Energies are given
  relative to the continuum edge $E_G$ in units of the phonon energy
  $\hbar\omega_{LO}$.  The temperature is 300 K. }
\label{fig:retardedGF}
\end{figure}
%%%%%%%%%%%%%%%%%%%%%%%%%%%%%%%%%%%%%%%%%%%%%%%%%%%%%%%%%%%%%%%%%%%%%%%%

The Fourier transform of the spectral function, -2 Im
$G^r_{\alpha}(\omega)$, is shown in Fig.~\ref{fig:retardedGF} for the
$k=0$ WL state and for the QD p- and s-shell (from top to bottom).  In
the absence of polar coupling to lattice distortions, the spectral
functions are delta functions at the free-particle energies indicated
by the vertical lines.  The dotted line in
Fig.~\ref{fig:retardedGF}(a) is the result for the $k=0$ WL state
without coupling to the QD states, indicating that their influence
on the WL polarons is weak.  The WL spectral function is
broadened, the central peak exhibits a small polaron shift, and
multiple sidebands due to LO-phonon emission (absorption) appear to
the right (left).  The polaron broadening is a result of the irreversible
decay in the continuous WL density of states.

The LO-phonon sidebands of the localized QD states in
Fig.~\ref{fig:retardedGF}(b) and (c) are more pronounced and the
hybridization of peaks from one shell with the energetically close
sidebands of the other shell can be observed.  This effect stems from
the discrete nature of the localized states and requires that the
coupling strength, which is modified by the form factors in
Eq.~(\ref{eq:coupling}), exceeds the polaron damping.  If only
coupling matrix elements diagonal in the state index would be
considered, a series of sidebands of a state with discrete energy,
spaced by the LO-phonon energy, would be obtained. Off-diagonal
coupling elements alone would lead to a hybridization of discrete
levels as, e.g., in quantum optics where instead of the phonon-field a
monochromatic light field coupled to a two-level system is considered
\cite{Inoshita:97}.  When the level splitting equals the LO-phonon
energy, in the limit of weak damping the splitting of each line is
determined by the carrier-phonon coupling strength.  Due to the
interplay of diagonal and off-diagonal interaction matrix elements,
the QD spectral functions in Fig.~\ref{fig:retardedGF}(b) and (c) show
a series of satellites, each of them reflecting the hybridization.
The asymmetry stems from the difference between level spacing (40 meV)
and LO-phonon energy (36 meV). The broadening of peaks stems mainly
from the coupling to the WL states.  A finite LO-phonon lifetime of 5
ps due to anharmonic interaction between phonons has been included in
the calculations.

%%%%%%%%%%%%%%%%%%%%%%%%%%%%%%%%%%%%%%%%%%%%%%%%%%%%%%%%%%%%%%%%%%%%%%%%%%
\section{Carrier kinetics of relaxation and capture processes}

In this section we study consequences of the renormalized
quasi-particle properties on the scattering processes.  Fermi's golden
rule, which has been frequently used in the past, describes only
transition rates from fully populated initial into empty final
states. Proper balancing between in- and out-scattering events,
weighted with the population $f$ of the initial states and the blocking
$1-f$ of the final states, leads to the kinetic equation
%%%%%%%%%%%%%%%%
\begin{align}
    & \frac{\partial}{\partial t} \: f_\alpha
     = \frac{2 \pi}{\hbar} \: 
        \sum_{\beta,\vec{q}} \:|M_{\beta \alpha}(\vec{q})|^2
\label{eq:Boltzmann}  \\
    & \hspace{0.0cm} 
    \times \big\{ \: (1 - f_\alpha) f_{\beta} \big[ 
     (1+n_{LO}) \: \delta( \varepsilon_\alpha - \varepsilon_{\beta} + \hbar\omega_{LO} ) 
\nonumber \\
    & \hspace{2.3cm} 
    +   n_{LO}  \: \delta( \varepsilon_\alpha - \varepsilon_{\beta} - \hbar\omega_{LO} )
                                       \big]
\nonumber \\
    &   \hspace{0.3cm}
         - f_\alpha (1-f_{\beta}) \big[ 
        n_{LO}  \: \delta( \varepsilon_\alpha - \varepsilon_{\beta} + \hbar\omega_{LO} ) 
\nonumber \\
    & \hspace{2.3cm} 
    +(1+n_{LO}) \: \delta( \varepsilon_\alpha - \varepsilon_{\beta} - \hbar\omega_{LO} )
                                       \big]
         \big\} \nonumber .
\end{align}
%%%%%%%%%%%%%%%
Quasiparticle and Markov approximation can also be applied to
renormalized polaronic states which results in a rate-equation
description for the population of these states \cite{Verzelen:00}.

A quantum-kinetic approach extends Eq.~(\ref{eq:Boltzmann}) in the
sense that the delta-functions with free-carrier energies are replaced
by time-integrals over polaronic retarded GFs. Furthermore, the
population factors are no longer instantaneous but explicitly depend
on the time evolution.  This is the time-domain picture for the inclusion of
renormalized quasi-particle properties (beyond a quasi-particle
approximation and beyond Markov approximation).  Using the generalized
Kadanoff-Baym ansatz (GKBA) \cite{Haug:}, the quantum-kinetic equation
has the form
%%%%%%%%%%%%%%%%
\begin{multline}
   \pdiff{}{t_1} f_{\alpha}(t_1) = 2 {\rm Re} \sum_{\beta} 
   \int_{-\infty}^{t_1} \! dt_2 
   ~~G^r_{\beta}(t_1-t_2) ~\big[G^r_{\alpha}(t_1-t_2)\big]^* \\
      \times \Big\{ \big[ 1-f_{\alpha}(t_2) \big] f_{\beta}(t_2)
                         ~i\hbar D^>_{\alpha \beta}(t_2-t_1) \\
                   - f_{\alpha}(t_2) \big[ 1-f_{\beta}(t_2) \big] 
                         ~i\hbar D^<_{\alpha \beta}(t_2-t_1)  \Big\} .
\label{eq:qkin}
\end{multline}
%%%%%%%%%%%%%%%%
The phonon propagator $D^>$ follows from Eq.~(\ref{eq:pn_propagator})
by replacing $\tau \rightarrow -\tau$.  A Markov approximation {\em in the
renormalized quasi-particle picture} corresponds to the assumption of a
slow time-dependence of the population $f_{\alpha}(t_2)$ in comparison
to the retarded GFs such that the population can be taken at the
external time $t_1$. The Boltzmann scattering integral of
Eq.~(\ref{eq:Boltzmann}) follows if one additionally neglects
quasi-particle renormalizations and uses free-carrier retarded GFs
\cite{note:FT}.

%%%%%%%%%%%%%%%%%%%%%%%%%%%%%%%%%%%%%%%%%%%%%%%%%%%%%%%%%%%%%%%%%%%%%%%%
\begin{figure}[htb]
\includegraphics*[width=.45\textwidth]{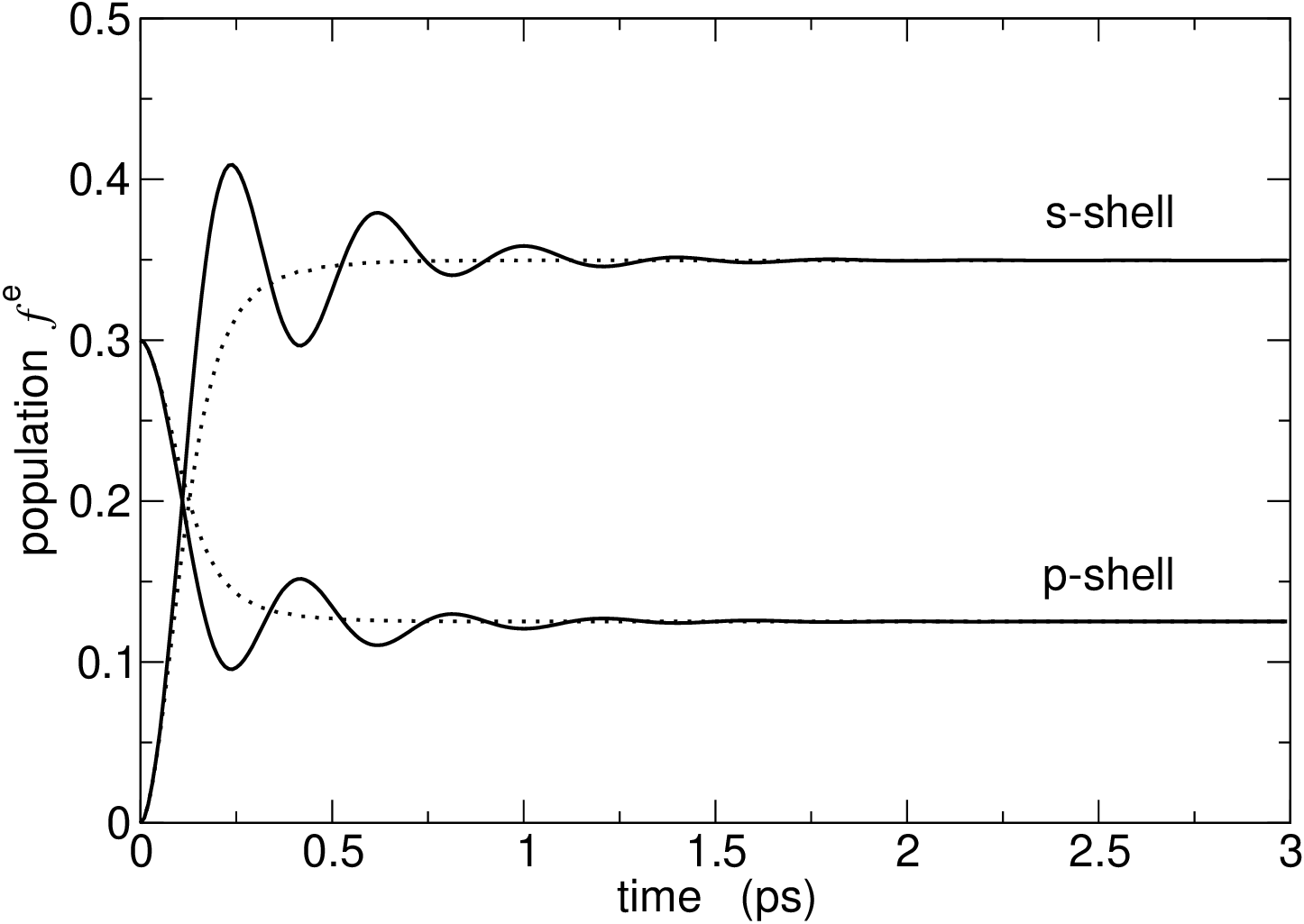}
\caption{Temporal evolution of the QD population due to carrier-phonon
         scattering between p-shell and s-shell for an energy spacing
         larger than the LO-phonon energy. The solid lines correspond to
         a quantum-kinetic calculation whereas for the dotted line
         the Markov approximation is used together with polaronic spectral
         functions.
         }
\label{fig:relaxation}
\end{figure}
%%%%%%%%%%%%%%%%%%%%%%%%%%%%%%%%%%%%%%%%%%%%%%%%%%%%%%%%%%%%%%%%%%%%%%%%

To demonstrate the influence of quantum-kinetic effects due to
QD-polarons, we first study the relaxation of carriers from p-shell to
s-shell for the above discussed situation where the level spacing does
not match the LO-phonon energy such that both, Fermi's golden rule and
the kinetic equation (\ref{eq:Boltzmann}) predict the absence of
transitions.  A direct time-domain calculation of the polaron GFs from
Eq.~(\ref{eq:Dyson_G_ret})-(\ref{eq:coupling}) together with
Eq.~(\ref{eq:qkin}) is used.  We assume an initial population
$f_s(t_0)=0$, $f_p(t_0)=0.3$ and start the calculation at time $t_0$.
While this example addresses the relaxation process itself, more
advanced calculations would also include the carrier generation via
optical excitation or carrier capture discussed below.  Then
ambiguities due to initial conditions can be avoided since the
population vanishes prior to the pump process which naturally provides
the lower limit of the time integral in Eq.~(\ref{eq:qkin}).  In
practice, we find that within the GKBA results weakly depend on the
details of the initial conditions.

The evaluation of the quantum-kinetic theory (solid lines in
Fig.~\ref{fig:relaxation}) yields a fast population increase of the
initially empty QD s-shell accompanied by oscillations which reflect
in the time-domain the hybridization of coupled carrier and phonon
states.  The analogy to Rabi oscillations has been pointed out in
Ref.~\cite{Inoshita:97}.  If one uses the Markov approximation
together with polaronic retarded GFs in Eq.~(\ref{eq:qkin}), such that
quasi-particle renormalizations are still included, these transient
oscillations disappear. In both cases the same steady-state solution
is obtained which corresponds to a thermal population at the
renormalized energies.  The equilibrium solution can be obtained from
the polaron spectral function using the Kubo-Martin-Schwinger (KMS)
relation,
%%%%%%%%%%%%%%%%
%\begin{equation}
$
  f_{\alpha}=-\int \frac{ d \hbar \omega }{\pi } ~~
  f(\omega) ~ {\rm Im} G_{\alpha}^r(\omega)
$
%\label{eq:KMS}
%\end{equation}
%%%%%%%%%%%%%%%
where $f(\omega)$ is a Fermi function with the lattice temperature.
Note that particle number conservation is obeyed in
Fig.~\ref{fig:relaxation} since the degeneracy of the p-shell is twice
that of the s-shell.

%%%%%%%%%%%%%%%%%%%%%%%%%%%%%%%%%%%%%%%%%%%%%%%%%%%%%%%%%%%%%%%%%%%%%%%%
\begin{figure}[htb]
\includegraphics*[width=.45\textwidth]{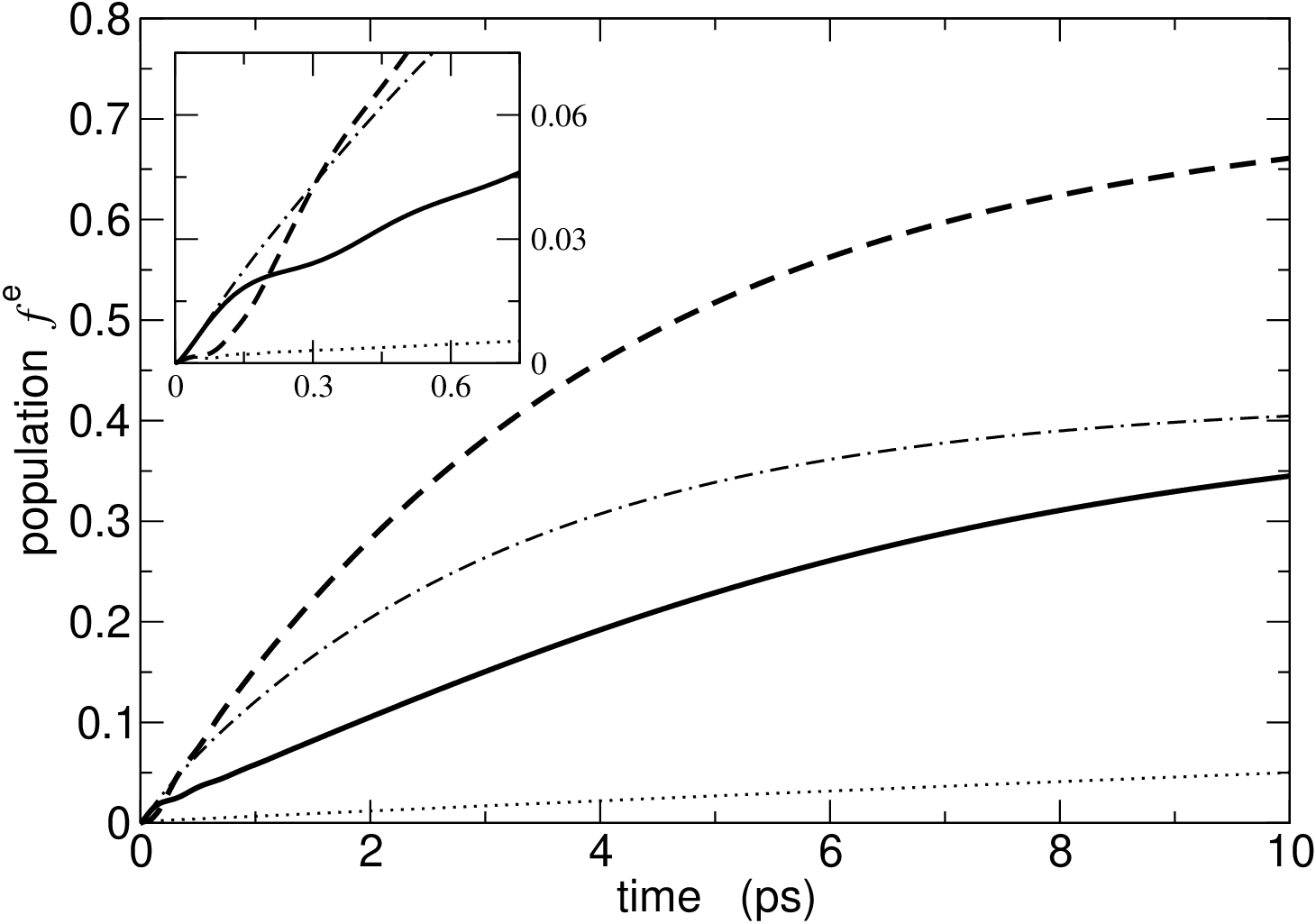}
\caption{Time-evolution of the QD p-shell (solid line) and s-shell 
         (dashed line) electron population due to carrier capture
         from the WL including the effect of carrier relaxation 
         between QD shells. If only direct capture processes are considered,
         the dashed-dotted and dotted lines are obtained for p-shell
         and s-shell, respectively.
        }
\label{fig:capture}
\end{figure}
%%%%%%%%%%%%%%%%%%%%%%%%%%%%%%%%%%%%%%%%%%%%%%%%%%%%%%%%%%%%%%%%%%%%%%%%

Another important process is the capture of carriers from the
delocalized WL states into the localized QD states. For the used QD
parameters, where the spacing between the p-shell and the lowest WL
state (40 meV) exceeds the LO-Phonon energy, again Fermi's golden rule
and Eq.~(\ref{eq:Boltzmann}) predict the absence of electronic
transitions. For the numerical solution of Eq.~(\ref{eq:qkin}) we use
now as initial condition empty QD states and a thermal population of
carriers in the polaronic WL states (obtained from the KMS relation)
corresponding to a carrier density 10$^{11}$ cm$^{-2}$ and temperature
300 K \cite{note:spec_func}.  The dashed-dotted and dotted lines in
Fig.~\ref{fig:capture} show the increase of the p- and s-shell
population, respectively, when only capture processes are considered
(scattering from a WL-polaron to a QD-polaron state due to emission of
LO-phonons).  Also in this situation the quantum-kinetic theory
predicts a fast population of the initially empty p-shell.  Albeit the
large detuning (exceeding two LO-phonon energies) the direct capture
to the s-shell is still possible but considerably slower.  When both,
direct capture of carriers as well as relaxation of carriers between
the QD states are included in the calculation, the solid (dashed) line
is obtained for the p-shell (s-shell) population. While faster capture
to the p-shell states leads at early times to a p-shell population
exceeding the s-shell populations (see inset of
Fig.~\ref{fig:capture}), the subsequent relaxation efficiently
populates the s-states. Since the WL states form a quasi-continuum,
beating at early times is strongly suppressed.

%%%%%%%%%%%%%%%%%%%%%%%%%%%%%%%%%%%%%%%%%%%%%%%%%%%%%%%%%%%%%%%%%%%%%%%%
\begin{figure}[htb]
\includegraphics*[width=.45\textwidth]{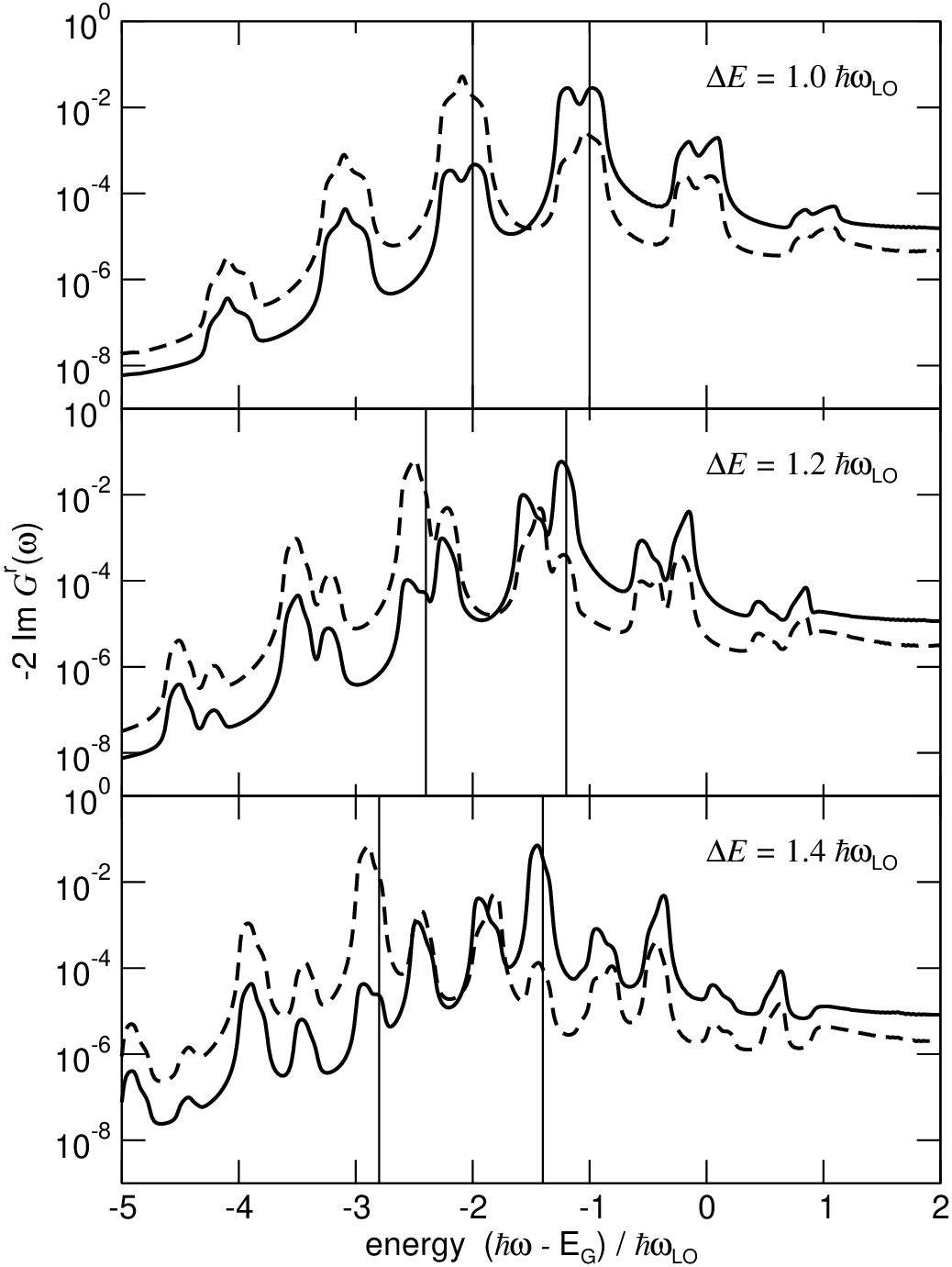}
\caption{(a) Spectral function of electrons in the coupled QD-WL
  system for the p-shell (solid line) and s-shell (dashed line) using
  various energy spacings $\Delta E$ between s-shell, p-shell 
  and WL ($k=0$) in units of the LO-phonon energy
  $\hbar\omega_{LO}$. Vertical lines indicate the positions of the 
  unrenormalized QD states, $E_G$ is the continuum edge of the WL states.
  The temperature is 300 K. }
\label{fig:retGFdetun}
\end{figure}
%%%%%%%%%%%%%%%%%%%%%%%%%%%%%%%%%%%%%%%%%%%%%%%%%%%%%%%%%%%%%%%%%%%%%%%%

%%%%%%%%%%%%%%%%%%%%%%%%%%%%%%%%%%%%%%%%%%%%%%%%%%%%%%%%%%%%%%%%%%%%%%%%
\begin{figure}[htb!]
\includegraphics*[width=.38\textwidth]{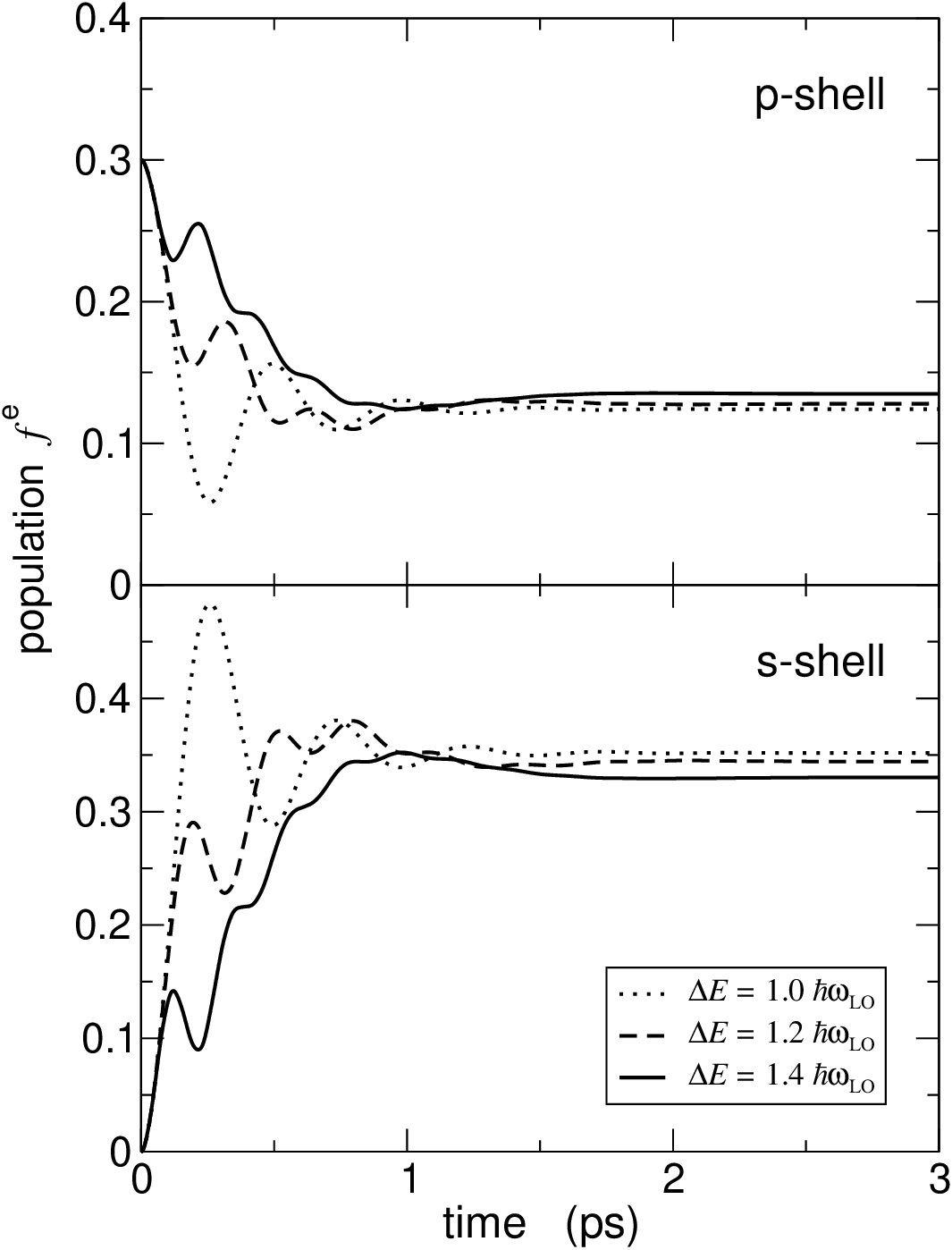}
\caption{Temporal evolution of the QD population due to carrier-phonon
         scattering between p-shell (initially populated) and s-shell
         (initially empty) for different energy spacings $\Delta E$
         corresponding to Fig.~\ref{fig:retGFdetun}.  }
\label{fig:relax_detun}
\end{figure}
%%%%%%%%%%%%%%%%%%%%%%%%%%%%%%%%%%%%%%%%%%%%%%%%%%%%%%%%%%%%%%%%%%%%%%%%

%%%%%%%%%%%%%%%%%%%%%%%%%%%%%%%%%%%%%%%%%%%%%%%%%%%%%%%%%%%%%%%%%%%%%%%%
\begin{figure}[htb!]
\includegraphics*[width=.38\textwidth]{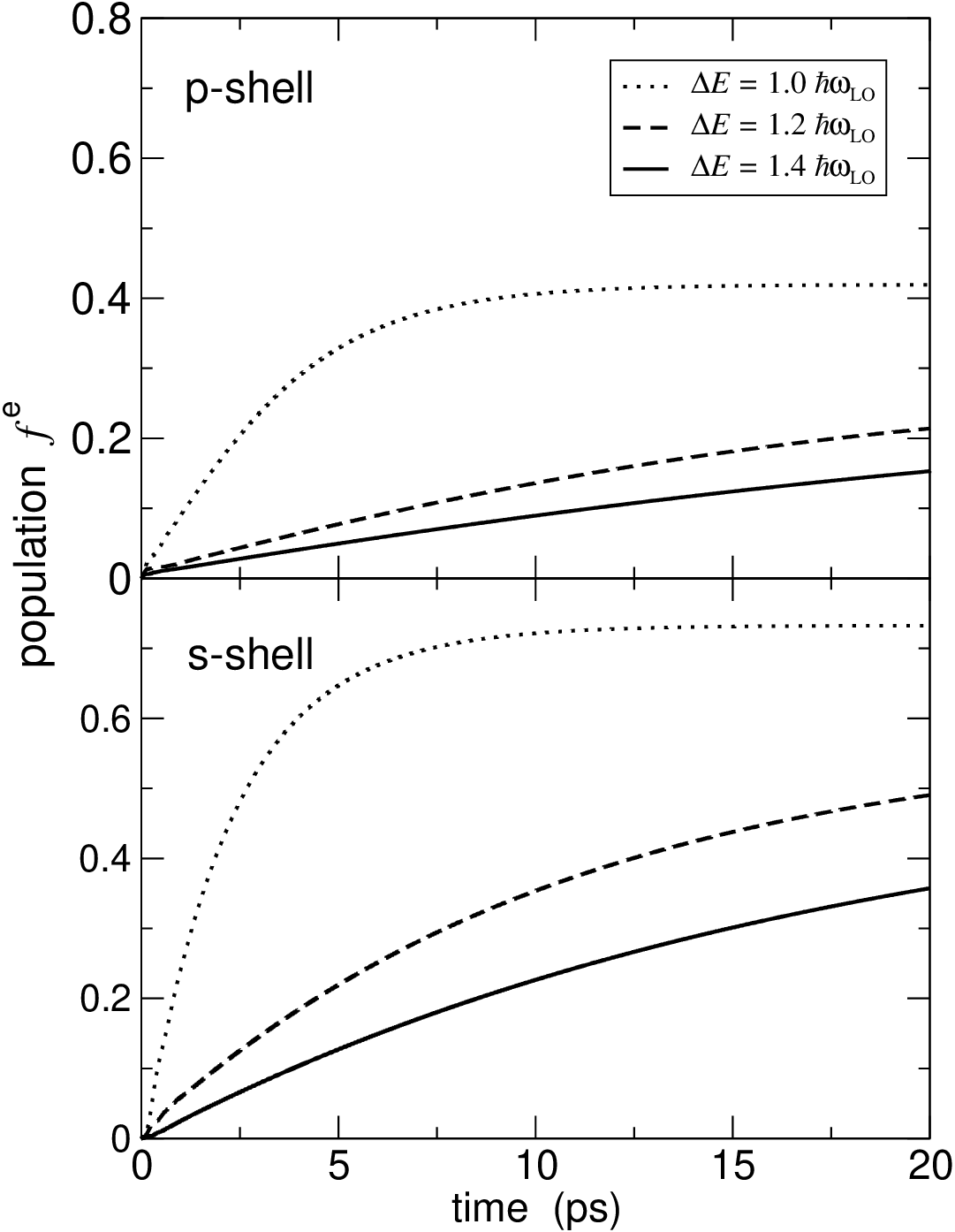}
\caption{Time-evolution of the QD p-shell (top) and s-shell 
         (bottom) electron population due to carrier capture
         from the WL including the effect of carrier relaxation 
         between QD shells for different energy spacings $\Delta E$
         corresponding to Fig.~\ref{fig:retGFdetun}. }
\label{fig:capt_detun}
\end{figure}
%%%%%%%%%%%%%%%%%%%%%%%%%%%%%%%%%%%%%%%%%%%%%%%%%%%%%%%%%%%%%%%%%%%%%%%%

With the results in Figs.~\ref{fig:retardedGF}-\ref{fig:capture} we
have demonstrated the ultrafast (subpicosecond) carrier relaxation and
fast (picosecond) carrier capture for a material with weak polar
coupling and 10\% detuning between the transition energies and the
LO-phonon energy.  This detuning is on the one hand sufficiently large
for the alternative LO+LA mechanism proposed by Inoshita and Sakaki
\cite{Inoshita:92} to fail and on the other hand small enough to
illustrate the hybridization of one state with sidebands of the other
states.  We find that the fast scattering is not related to the near
resonance condition and in fact relatively insensitive to the detuning
between transition energies and LO-phonon energy. The spectral
functions of the coupled QD-WL system for various detunings, ranging
from resonance to a large mismatch of 40\%, are shown in
Fig.~\ref{fig:retGFdetun}. For better visibility only the curves for
the s- and p-shells are displayed, while the WL-states are also
included in the calculations. As seen in Fig.~\ref{fig:retardedGF},
the spectral function for the WL states is only weakly influenced by
the coupling to the QDs.  In all three cases of
Fig.~\ref{fig:retGFdetun} there is substantial overlap between the
s-shell and p-shell density of states which points to efficient
transition processes. This overlap is due to the multi-peak-structure
which contains the series of phonon sidebands spaced by the LO-phonon
energy and their hybridization. From top to bottom in
Fig.~\ref{fig:retGFdetun}, the peak splitting increases with detuning.

The corresponding results for the carrier relaxation, as in
Fig.~\ref{fig:relaxation} but for different detunings $\Delta E$, are
shown in Fig.~\ref{fig:relax_detun}. The fast carrier relaxation
towards an equilibrium situation is retained in all three cases. The
main difference is in the oscillation period, which is reduced for
larger detuning due to the increased splitting in
Fig.~\ref{fig:retGFdetun}.

A stronger influence of the detuning between transition energies and
the LO-phonon energy is found for the capture of carriers from the WL
into the QD states. As can be seen in Fig.~\ref{fig:capt_detun}, from
the resonance situation to a detuning of 40\% the capture efficiency
is reduced by about one order of magnitude. Nevertheless, a
significant occupancy can be reached within several ten picoseconds.
The reduced capture efficiency is related to a reduced overlap between
the WL and QD spectral functions for increasing detuning (which is
mainly because the WL states are weakly influenced by the QD states).
In comparison to this, the strong interaction between s- and p-states
maintains a strong overlap between their spectral functions. As a
consequence the relaxation time is less sensitive to the detuning.

In summary, the quantum-kinetic treatment of carrier-phonon
interaction explains the absence of a phonon bottleneck in terms of
scattering between renormalized quasi-particle states. A
quasi-equilibrium situation is reached on a ps-timescale at elevated
temperatures even in materials with weak polar coupling. 

\begin{acknowledgments}
  This work was supported by the Deutsche Forschungsgemeinschaft and
  with a grant for CPU time at the NIC, Forschungszentrum J\"{u}lich. 
\end{acknowledgments}

\newpage

%%%%%%%%%%%%%%%%%%%%%%%%%%%%%%%%%%%%%%%%%%%%%%%%%%%


\begin{thebibliography}{10}
\bibitem{note:diag_GF} In general, Green's functions and self-energies
  depend on two quantum numbers $\alpha,\beta$. For the situation described 
  in this paper, off-diagonal contributions can be neglected, which has been 
  verified numerically; see also \cite{Inoshita:97,Kral:98} .
\bibitem{note:RPA} Comparison with an alternative method supports this
  statement: the first term of a cumulant expansion, which is known to
  be exact for the independent Boson model \cite{Mahan:90}, agrees well
  with the RPA for quasi-continuous electronic states.
\bibitem{note:FT} The quantum-kinetic equation provides
  energy-conserving scattering processes only in the long-time limit.
  The corresponding spectral function in Fig.~\ref{fig:retardedGF} is
  obtained from a complete Fourier transform whereas on a short
  timescale the peaks shown there are only partially emerged (since
  only a finite-time contribution of the retarded GF enters in the
  quantum-kinetic equation at a given time.)
\end{thebibliography}
\end{document}